\documentclass[%
 reprint,
 amsmath,amssymb,
 aps,
]{revtex4-1}

\usepackage{graphicx}
\usepackage{dcolumn}
\usepackage{bm}

\begin{document}

\preprint{APS/123-QED}
\title{Solution of two level system using $3^{rd}$ order Magnus expansion}
\thanks{the author thanks Prof. Raphael D Levine of huji and Dr. Yonatan Dubi of BGU}%

\author{Dawit Hiluf Hailu}
\email{dawit@post.bgu.ac.il}
\affiliation{Ben-Gurion University of Negev department of Physical Chemistry, Be'er-Sheva 84105, Israel}
\affiliation{Physics Department, Mekelle University, P.O.Box 231, Mekelle, Ethiopia.}
\date{\today}
\begin{abstract}
The solution of a two level system driven by a Laser in the adiabatic limit is determined using third order Magnus expansion. We made the assumption that the laser is on resonance or close to resonance with the Bohr transition. As a consequence of which we are able to obtain a Hamiltonian which commute with itself at different times. We solve the problem using  the Sylvester Formula where we make use of the eigenvalues.
\end{abstract}

\maketitle


\section{Introduction}
The study of the dynamics of two level system has been attracting interest of researchers in different areas ranging from Nuclear magnetic resonance (NMR) \cite{Eberly1975} to Quantum Computers \cite{MichaelChuang2010}. The solution of a two level system whose Hamiltonian is closed under Lie algebra can be obtained using Wei-Norman \cite{wei1963lie} and is pursued by different authors \cite{dattoli1988evolution, altafiniuse}. As application the author in \cite {altafiniuse}, for example, has used the solution obtained via Wei-Norman for quantum computing. The solution obtained via Wei-Norman is exact, yet it is sometimes insightful to get analytical solutions of the dynamics of the system. We here aim to use the knowledge of the eigenvalues to provide analytical solution of two level system using the third order Magnus expansion via Sylvester formula. Moreover the focus in this paper is for an isolate system, however, if we include noise the coherences will be destroyed by the effect of the environment.

The outline of the paper is as follows: in section (\ref{sec:thesystem}) we introduce the two level system where we describe its Hamiltonian and obtain the equation of motion, in section (\ref{sec:Magnus}) we use third order Magnus expansion to obtain the solution of the equation of motion, in section (\ref{sec:results}) we present the comparison of the solutions of the equation of motion using the third order Magnus expansion and numerical solution. We will also present comparison between first, second and third order Magnus expansions. At last in section (\ref{sec:conclusion}) we summarize our results and provide an outlook. 
\section{The system}
\label{sec:thesystem}
\begin{figure}[htbp]
\begin{center}
\includegraphics[width=2.0 in]{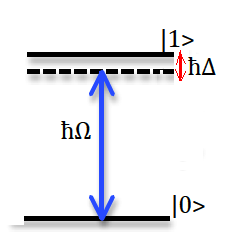}
\caption{(Color online) Two level system. Detuning $\hbar\Delta$ and coupling $\hbar\Omega\left(t\right)$ }
\label{fig2}
\end{center}
\end{figure}

The system we consider is shown in fig.(\ref{fig2}), we assume to  have a two level system with ground state $|0\rangle$ and excited stats $|1\rangle$. Suppose the two levels are coupled by a field $\Omega \left(t\right)$ and the laser is off by detuning $\Delta$, which is the difference between the laser frequency and the Bohr frequency. We consider an adiabatic population transfer \cite{Bergmann2001}, yet we apply a weak pulse (corresponding to pulse area of $\pi/2$) to avoid transferring all the population, from $|0\rangle$ to $|1\rangle$, as one would have when using pulse area of $\pi$. Using $\pi$ pulse area protocol one could completely transfer the populations from the ground state $|0\rangle$ to the excited stats $|1\rangle$ \cite{Vitanov:2001aa}.
\graphicspath{{Figures//}}

The Hamiltonian under Rotating Wave Approximation (RWA) and in the interaction picture is known to be \cite{Bergmann2001,shore2008}.
\begin{equation}
\hat H \left(t\right)=\frac{\hbar}{2}\begin{pmatrix}
0 & \Omega  \left(t\right)\\
\Omega \left(t\right) & 2\Delta \\
 \end{pmatrix} 
\end{equation}
 We use the Pauli matrices as our generators, and they are given by \cite{MichaelChuang2010}
\begin{equation}
\begin{aligned}
 \hat \sigma_1=\begin{pmatrix}
0 & 1\\
1 & 0 \end{pmatrix},       & & \hat \sigma_2=\begin{pmatrix}
0 & -i\\
i & 0 \end{pmatrix} ,      & & \hat \sigma_3=\begin{pmatrix}
1 & 0\\
0 & -1 
 \end{pmatrix}  
 \end{aligned}
\end{equation}
In terms of which we expand the density matrix as well as the Hamiltonian \cite{Hioe1981}.
 \begin{equation}
\begin{aligned}
\hat\rho\left(t\right)=&\frac{\hat I}{2}+\frac{1}{2}\sum_{j=1}^{3} S_j\left(t\right)\hat \sigma_j\\
\hat H\left(t\right)=&\frac{\hbar}{2}\left[\left(\sum_{k}^2\omega_k\right)\hat I+\sum_{j=1}^{3} \gamma_j\left(t\right)\hat \sigma_j\right]
\end{aligned}
\end{equation}
where $\hbar\omega_k$ is energy of level $k$ and $\hat I$ is the identity operator. The coefficients $S_j\left(t\right)$ and $\gamma_j\left(t\right)$ are given by \cite{Hioe1981}
\begin{subequations}
\begin{align}
S_j\left(t\right)=&Tr\left(\hat\rho\left(t\right)\hat \sigma_j\right)\label{expectS}\\
\hbar\gamma_j\left(t\right)=&Tr\left(\hat H\left(t\right)\hat \sigma_j\right)\label{Torque}
\end{align}
\end{subequations}
the generators $\hat \sigma_j$ are the Pauli matrices and has the following properties \cite{Hioe1981}
\begin{equation}
\begin{aligned}
Tr\left(\hat \sigma_i\hat \sigma_j\right)=& 2\delta_{ij}\\
\left[\hat \sigma_{i}, \hat \sigma_{j}\right]=&2i \epsilon_{ijk} \hat \sigma_k
\end{aligned}
\end{equation}
Based on this, and following the same procedure as in \cite{Hioe1981,AlhassidLevine} we get the  equation of motion for the coherence vector. If we now form vector $\vec S=\left(S_1\left(t\right),S_2\left(t\right),S_3\left(t\right)\right)^T$ whose elements are the expectation value of the Pauli matrices as given by Eq.\eqref{expectS}, we readily obtain the equation of motion to be
\begin{equation}
\begin{aligned}
\frac{d}{dt}\vec{S}=&g\vec S
\label{sdot}
\end{aligned}
\end{equation}
 where $g$ is given by 
\begin{equation}
\begin{aligned}
g=&
 \begin{pmatrix}
  0 & \Delta  & 0 \\
  -\Delta & 0 & -\Omega\left(t\right) \\
  0 & \Omega\left(t\right) & 0
 \end{pmatrix}
\end{aligned}
\label{matrixg}
\end{equation}
Although one can solve this equation of motion numerically, we are here interested in finding a comparison between numerical and analytical solution. We will solve Eq.\eqref{sdot} both numerically and analytically, we then compare the  two solutions for different values of detuning.  
\section{Magnus Approximations}
\label{sec:Magnus}
One way of solving equation Eq.\eqref{sdot} is given by the Magnus Approximation \cite{Magnus1954}.The Magnus expansion, named after Wilhem Magnus, provides an exponential representation of the solution of a first order linear homogenous equation for linear operator. Given an $N\times N$ coefficient matrix $g\left(t\right)$ we wanted to solve the initial value value problem associated with the linear ordinary differential equation, which in our case is the equation of motion for the coherence vectors given by, along with its initial condition:
\begin{equation}
\begin{aligned}
\frac{d \vec{S}\left(t\right)}{dt}=&g\left(t\right)\vec S \left(t\right) &     \vec S \left(0\right) =S_0
\end{aligned}
\end{equation}
We now make the assumption that the laser is on resonance or close to resonance, meaning $\Delta=0$ or is negligibly small and  consequently the matrix $g$ now commutes with itself at different times, that is $\left[g\left(t_1\right),g\left(t_2\right)\right]=0$. 
The approach proposed by Magnus to solve the matrix initial value problem is to express the solution of the exponential of a certain $N\times N$ function $G\left(t,t_0\right)$
\begin{equation}
\begin{aligned}
\vec{S}\left(t\right)=&e^{G\left(t,0\right) }\vec S \left(0\right)
\end{aligned}
\end{equation}
which is subsequently written as a series expansion
\begin{equation}
\begin{aligned}
G\left(t,0\right)=&\sum_{k=1}^\infty G_k\left(t,0\right)
\end{aligned}
\end{equation}
writing $G\left(t,0\right)=G\left(t\right)$ for simplicity, the first three series reads thus 
 \begin{equation}
\begin{aligned}
G_1\left(t\right)=&\int_{0}^t g\left(t_1\right) dt_1\\
G_2\left(t\right)=&\frac{1}{2}\int_{0}^t dt_1 \int_{0}^{t_1} \left[g\left(t_1\right), g\left(t_2\right)\right] dt_2\\
G_3\left(t\right)=&\frac{1}{6}\int_{0}^t dt_1 \int_{0}^{t_1} dt_2 \int_{0}^{t_2} \left[g\left(t_1\right),\left[g\left(t_2\right), g\left(t_3\right)\right]\right]dt_3\\
+&\frac{1}{6}\int_{0}^t dt_1 \int_{0}^{t_1} dt_2 \int_{0}^{t_2}\left [g\left(t_3\right),\left[g\left(t_2\right), g\left(t_1\right)\right]\right]  dt_3
\end{aligned}
\label{Gseries}
\end{equation}
where $\left[g_1,g_2\right]=g_1g_2-g_2g_1$.  In search of better approximation one needs to include more terms in the Magnus expansion. In this paper we will be considering only  the first three series $G_k\left(t\right), k=1, 2, 3 $, therefore the solution for Eq.\eqref{sdot} now becomes
\begin{equation}
\begin{aligned}
\vec{S}\left(t\right)=&e^{G^{(3)}\left(t\right)}\vec S \left(0\right)
\end{aligned}
\label{Soft}
\end{equation}
where $G^{(3)}\left(t\right)$ is sum of the three series given in Eq.\eqref{Gseries}. One way of determining $e^{G^{(3)}}$ is using the Sylvester formula. The Sylvester formula is a way of solving any exponential function by making use of eigenvalues \cite{19dubiousways,Tarantola}. To this end let $\gamma_j$  be an eigenvalue of $G^{(3)}\left(t\right)$, we thus can write the exponent using Sylvester formula (iff we have distinct eigenvalues) as  
\begin{equation}
\begin{aligned}
e^{G^{(3)}\left(t\right)}=&\sum_{j=1}^{3}e^{\gamma_j}\prod_{j\neq k=1}^{3}\frac{G^{(3)}\left(t\right)-\gamma_j  I}{\gamma_k-\gamma_j}
\end{aligned}
\label{sylvstr}
\end{equation}
The third order Magnus is  obtained to be of the form  
\begin{equation}
\begin{aligned}
G^{(3)}\left(t\right)=&\begin{pmatrix}
 0 &  \eta\left(t\right)  & \lambda\left(t\right) \\
  -\eta\left(t\right) & 0 & -\zeta\left(t\right) \\
  -\lambda\left(t\right) &  \zeta\left(t\right) & 0
 \end{pmatrix}
\end{aligned}
\label{Hintg}
\end{equation}
where $\eta\left(t\right)=\Delta'\left(t\right)+\lambda_1\left(t\right)$ and $\zeta\left(t\right)=\Omega'\left(t\right)+\lambda_2\left(t\right)$ with $\Omega' \left(t\right)=\int_{0}^t \Omega\left(t_1\right) dt_1$, $\Delta'\left(t\right)=\int_{0}^t \Delta  dt_1$ and
\begin{equation}
\begin{aligned}
\lambda\left(t\right)=&\frac{\Delta}{2}\int_{0}^t dt_1\int_{0}^{t_1} dt_2\left(\Omega_1-\Omega_2\right)\\
\lambda_1\left(t\right)=&-\frac{\Delta}{6}\int_{0}^t dt_1 \int_{0}^{t_1} dt_2 \int_{0}^{t_2} dt_3\left(\Omega_1\left(\Omega_2-2\Omega_3\right)+\Omega_2\Omega_3\right)\\
\lambda_2\left(t\right)=&-\frac{\Delta^2}{6}\int_{0}^t dt_1 \int_{0}^{t_1} dt_2 \int_{0}^{t_2} dt_3\left(\Omega_1-2\Omega_2+\Omega_3\right)
\end{aligned}
\label{integrands}
\end{equation}
where we used the notation $\Omega_j=\Omega\left(t_j\right)$ for simplicity.
In what follows  we omit the time argument $\left(t\right)$ unless it is needed for clarity. The eigenvalues of $G^{(3)}$, are readily obtained to be $\{0, -i\xi, i\xi\}$, where $\xi=\sqrt{\lambda^2+\zeta^2+\eta^2}$. It is worth pointing out here that we have distinct eigenvalues. 
Therefore we can now express our solution for the coherence vector in terms of the Sylvester formula. To this end, making use of the eigenvalues and  Eq.\eqref{Soft} along with Eq.\eqref{sylvstr}  and noting that the system is initially prepared to be on the ground state,  we find the following solution 
\begin{equation}
\begin{aligned}
\vec{S}\left(t\right)=&
 \begin{pmatrix}
 -\frac{\zeta\eta-\zeta\eta\cos{\xi}-\lambda\xi\sin{\xi}}{\xi^2}\\
  \frac{-\lambda\eta+\lambda\eta\cos{\xi}-\zeta\xi\sin{\xi}}{\xi^2}\\
 \frac{\eta^2+(\lambda^2+\xi^2)\cos{\xi}}{\xi^2}
 \end{pmatrix}
\end{aligned}
\label{st1}
\end{equation}
If however we consider only the first term in the Magnus expansion, it follows that $\lambda=\lambda_1=\lambda_2=0$, and  accordingly we have to modify our solution-- Eq.\eqref{st1} to be
\begin{equation}
\begin{aligned}
\vec{S}\left(t\right)=&
 \begin{pmatrix}
 \frac{\Delta'\Omega'}{\xi^2}\big(-1+\cos\xi\big)\\
 - \frac{\Omega'}{\xi}\sin\xi\\
 \frac{\Delta'^2}{\xi^2}+\frac{\Omega'^2}{\xi^2}\cos\xi
 \end{pmatrix}
\end{aligned}
\label{st2}
\end{equation}
where now $\xi=\sqrt{\Delta'^2+\Omega'^2}$
\section{Results and discussion}
\label{sec:results}
\begin{figure}[htbp]
\centering
\includegraphics[width=2.5 in]{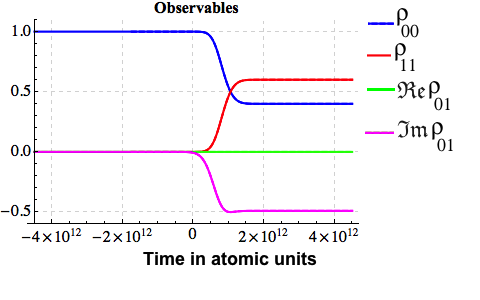}
\caption{(Color online) Solution using first order Magnus, i.e. when $\lambda=\lambda_1=\lambda_2=0$ and $\Delta=0$. The solid lines are numerical solution whereas the dashed lines are analytical solution. Key: blue=$\rho_{00}$, red=$\rho_{11}$, green=Real part of $\rho_{01}$, and magenta=Imaginary part of $\rho_{01}$ }
\label{Soln123ndOMagnusD0}
\end{figure}

\begin{figure}[htbp]
\centering
\includegraphics[width=4.5 in]{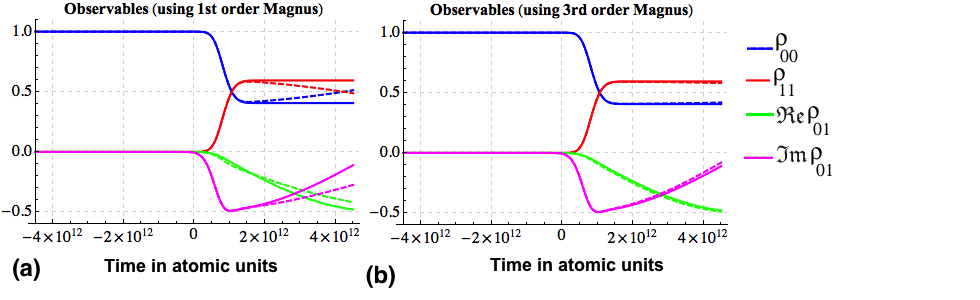}
\caption{(Color online) Solution using (a) first order Magnus, i.e. when $\lambda=\lambda_1=\lambda_2=0$~(b) third order Magnus, i.e. when $\lambda=\lambda_1=\lambda_2\neq0$. The solid lines are numerical solution whereas the dashed lines are analytical solution. Key: blue=$\rho_{00}$, red=$\rho_{11}$, green=Real part of $\rho_{01}$, and magenta=Imaginary part of $\rho_{01}$ }
\label{Soln3rdOMagnus}
\end{figure}

It is worth pointing out here that because of the pulse area we used, i.e. $\frac{\pi}{2}$ we do not see the Rabi oscillations, we instead see that, at the end of the pulse interaction, we are able to create  superpositions between states $|0\rangle$ and $|1\rangle$. To see the Rabi oscillation, where in the population fluctuates between states $|0\rangle$ and $|1\rangle$, one needs to use a laser whose pulse area is an integral multiple of $\pi$.  
For exact resonance where $\Delta=0$, as can be seen in Fig.(\ref{Soln123ndOMagnusD0}), the solutions obtained using Magnus expansion and numerical solution are identical. This is because all the commutator terms are dependent on $\Delta$. But if we have a non-zero detuning but small value the results obtained numerically and using Magnus expansion may differ. We see in Fig.(\ref{Soln3rdOMagnus}a) comparison of solutions obtained via numerical and first order Magnus. As we  took a very  small value of detuning  we have a very small value resulted from the commutation of the Hamiltonian with itself at different times. Consequently the solution we obtained using the first order Magnus approximation is close to the numerical but does not fully agree with the numerical solution. 
\graphicspath{{Figures//}}
Whereas if we use the third order Magnus expansion as it means we include more correction terms the  the solution is seen to be improved Fig.(\ref{Soln3rdOMagnus}b)
\graphicspath{{Figures//}}

\section{Conclusion}
\label{sec:conclusion}
In conclusion we have shown in this paper that for a two level system which is either on resonance or close to resonance one can obtain an analytical solution of the dynamics which is very much close to the exact solution obtained via numerical.  If the system is far from resonance case we recommend to include more of the commutation terms of the exponent to obtain a better approximate of the solution. To reiterate, for exact resonance where $\Delta=0$ the solutions obtained using Magnus expansion and numerical solution are identical. This is because all the commutator terms are dependent on $\Delta$. But if we have a non-zero detuning but small  value the results obtained numerically and using 3$^rd$ order Magnus expansion agrees quite well. If however the value of the detuning is big then the solutions differ. We plan to extend the  approach for three-level system in the SU(3) dynamics. One point worth mentioning is that the system we have considered is closed system but if we include a noise the coherence would be affected negatively.



\bibliography{mybibfile}

\begin{thebibliography}{13}%
\makeatletter
\providecommand \@ifxundefined [1]{%
 \@ifx{#1\undefined}
}%
\providecommand \@ifnum [1]{%
 \ifnum #1\expandafter \@firstoftwo
 \else \expandafter \@secondoftwo
 \fi
}%
\providecommand \@ifx [1]{%
 \ifx #1\expandafter \@firstoftwo
 \else \expandafter \@secondoftwo
 \fi
}%
\providecommand \natexlab [1]{#1}%
\providecommand \enquote  [1]{``#1''}%
\providecommand \bibnamefont  [1]{#1}%
\providecommand \bibfnamefont [1]{#1}%
\providecommand \citenamefont [1]{#1}%
\providecommand \href@noop [0]{\@secondoftwo}%
\providecommand \href [0]{\begingroup \@sanitize@url \@href}%
\providecommand \@href[1]{\@@startlink{#1}\@@href}%
\providecommand \@@href[1]{\endgroup#1\@@endlink}%
\providecommand \@sanitize@url [0]{\catcode `\\12\catcode `\$12\catcode
  `\&12\catcode `\#12\catcode `\^12\catcode `\_12\catcode `\%12\relax}%
\providecommand \@@startlink[1]{}%
\providecommand \@@endlink[0]{}%
\providecommand \url  [0]{\begingroup\@sanitize@url \@url }%
\providecommand \@url [1]{\endgroup\@href {#1}{\urlprefix }}%
\providecommand \urlprefix  [0]{URL }%
\providecommand \Eprint [0]{\href }%
\providecommand \doibase [0]{http://dx.doi.org/}%
\providecommand \selectlanguage [0]{\@gobble}%
\providecommand \bibinfo  [0]{\@secondoftwo}%
\providecommand \bibfield  [0]{\@secondoftwo}%
\providecommand \translation [1]{[#1]}%
\providecommand \BibitemOpen [0]{}%
\providecommand \bibitemStop [0]{}%
\providecommand \bibitemNoStop [0]{.\EOS\space}%
\providecommand \EOS [0]{\spacefactor3000\relax}%
\providecommand \BibitemShut  [1]{\csname bibitem#1\endcsname}%
\let\auto@bib@innerbib\@empty
\bibitem [{\citenamefont {Allen}\ and\ \citenamefont
  {Eberly}(1975)}]{Eberly1975}%
  \BibitemOpen
  \bibfield  {author} {\bibinfo {author} {\bibfnamefont {L.}~\bibnamefont
  {Allen}}\ and\ \bibinfo {author} {\bibfnamefont {J.~H.}\ \bibnamefont
  {Eberly}},\ }\href@noop {} {\emph {\bibinfo {title} {Optical resonance and
  two-level atoms}}}\ (\bibinfo  {publisher} {Wiley},\ \bibinfo {year}
  {1975})\BibitemShut {NoStop}%
\bibitem [{\citenamefont {Nielsen}\ and\ \citenamefont
  {Chuang}(2010)}]{MichaelChuang2010}%
  \BibitemOpen
  \bibfield  {author} {\bibinfo {author} {\bibfnamefont {M.~A.}\ \bibnamefont
  {Nielsen}}\ and\ \bibinfo {author} {\bibfnamefont {I.~L.}\ \bibnamefont
  {Chuang}},\ }\href@noop {} {\emph {\bibinfo {title} {Quantum computation and
  quantum information}}},\ \bibinfo {edition} {10th}\ ed.\ (\bibinfo
  {publisher} {Cambridge University Press},\ \bibinfo {year}
  {2010})\BibitemShut {NoStop}%
\bibitem [{\citenamefont {Wei}\ and\ \citenamefont
  {Norman}(1963)}]{wei1963lie}%
  \BibitemOpen
  \bibfield  {author} {\bibinfo {author} {\bibfnamefont {J.}~\bibnamefont
  {Wei}}\ and\ \bibinfo {author} {\bibfnamefont {E.}~\bibnamefont {Norman}},\
  }\href@noop {} {\bibfield  {journal} {\bibinfo  {journal} {Journal of
  Mathematical Physics}\ }\textbf {\bibinfo {volume} {4}},\ \bibinfo {pages}
  {575} (\bibinfo {year} {1963})}\BibitemShut {NoStop}%
\bibitem [{\citenamefont {Dattoli}\ \emph {et~al.}(1988)\citenamefont
  {Dattoli}, \citenamefont {Richetta},\ and\ \citenamefont
  {Torre}}]{dattoli1988evolution}%
  \BibitemOpen
  \bibfield  {author} {\bibinfo {author} {\bibfnamefont {G.}~\bibnamefont
  {Dattoli}}, \bibinfo {author} {\bibfnamefont {M.}~\bibnamefont {Richetta}}, \
  and\ \bibinfo {author} {\bibfnamefont {A.}~\bibnamefont {Torre}},\
  }\href@noop {} {\bibfield  {journal} {\bibinfo  {journal} {Journal of
  mathematical physics}\ }\textbf {\bibinfo {volume} {29}},\ \bibinfo {pages}
  {2586} (\bibinfo {year} {1988})}\BibitemShut {NoStop}%
\bibitem [{\citenamefont {Claudio}()}]{altafiniuse}%
  \BibitemOpen
  \bibfield  {author} {\bibinfo {author} {\bibfnamefont {A.}~\bibnamefont
  {Claudio}},\ }in\ \href@noop {} {\emph {\bibinfo {booktitle} {15th
  International Symposium on Mathematical Theory of Networks and Systems, Notre
  Dame, IN, USA, University of Notre Dame}}},\ pp.\ \bibinfo {pages}
  {12--16}\BibitemShut {NoStop}%
\bibitem [{\citenamefont {Viatnov}\ \emph {et~al.}(2001)\citenamefont
  {Viatnov}, \citenamefont {Halfmann}, \citenamefont {Shore},\ and\
  \citenamefont {Bergmann}}]{Bergmann2001}%
  \BibitemOpen
  \bibfield  {author} {\bibinfo {author} {\bibfnamefont {N.~V.}\ \bibnamefont
  {Viatnov}}, \bibinfo {author} {\bibfnamefont {T.}~\bibnamefont {Halfmann}},
  \bibinfo {author} {\bibfnamefont {B.~W.}\ \bibnamefont {Shore}}, \ and\
  \bibinfo {author} {\bibfnamefont {K.}~\bibnamefont {Bergmann}},\ }\href@noop
  {} {\bibfield  {journal} {\bibinfo  {journal} {Annual Review of Physical
  Chemistry}\ }\textbf {\bibinfo {volume} {52}},\ \bibinfo {pages} {763}
  (\bibinfo {year} {2001})}\BibitemShut {NoStop}%
\bibitem [{\citenamefont {Vitanov}\ \emph {et~al.}(2001)\citenamefont
  {Vitanov}, \citenamefont {Halfmann}, \citenamefont {Shore},\ and\
  \citenamefont {Bergmann}}]{Vitanov:2001aa}%
  \BibitemOpen
  \bibfield  {author} {\bibinfo {author} {\bibfnamefont {N.~V.}\ \bibnamefont
  {Vitanov}}, \bibinfo {author} {\bibfnamefont {T.}~\bibnamefont {Halfmann}},
  \bibinfo {author} {\bibfnamefont {B.~W.}\ \bibnamefont {Shore}}, \ and\
  \bibinfo {author} {\bibfnamefont {K.}~\bibnamefont {Bergmann}},\ }\href@noop
  {} {\bibfield  {journal} {\bibinfo  {journal} {Annu. Rev. Phys. Chem.}\
  }\textbf {\bibinfo {volume} {52}},\ \bibinfo {pages} {763} (\bibinfo {year}
  {2001})}\BibitemShut {NoStop}%
\bibitem [{\citenamefont {Shore}(2008)}]{shore2008}%
  \BibitemOpen
  \bibfield  {author} {\bibinfo {author} {\bibfnamefont {B.~W.}\ \bibnamefont
  {Shore}},\ }\href@noop {} {\bibfield  {journal} {\bibinfo  {journal} {Acta
  Physica Slovaca Reviews and Turorials}\ }\textbf {\bibinfo {volume} {58}},\
  \bibinfo {pages} {243} (\bibinfo {year} {2008})}\BibitemShut {NoStop}%
\bibitem [{\citenamefont {Hioe}\ and\ \citenamefont {Eberly}(1981)}]{Hioe1981}%
  \BibitemOpen
  \bibfield  {author} {\bibinfo {author} {\bibfnamefont {F.~T.}\ \bibnamefont
  {Hioe}}\ and\ \bibinfo {author} {\bibfnamefont {J.~H.}\ \bibnamefont
  {Eberly}},\ }\href@noop {} {\bibfield  {journal} {\bibinfo  {journal} {Phys.
  Rev. Lett.}\ }\textbf {\bibinfo {volume} {47}},\ \bibinfo {pages} {838}
  (\bibinfo {year} {1981})}\BibitemShut {NoStop}%
\bibitem [{\citenamefont {Alhassid}\ and\ \citenamefont
  {Levine}(1978)}]{AlhassidLevine}%
  \BibitemOpen
  \bibfield  {author} {\bibinfo {author} {\bibfnamefont {Y.}~\bibnamefont
  {Alhassid}}\ and\ \bibinfo {author} {\bibfnamefont {R.~D.}\ \bibnamefont
  {Levine}},\ }\href@noop {} {\bibfield  {journal} {\bibinfo  {journal} {Phys.
  Rev. A}\ }\textbf {\bibinfo {volume} {18}},\ \bibinfo {pages} {89} (\bibinfo
  {year} {1978})}\BibitemShut {NoStop}%
\bibitem [{\citenamefont {Magnus}(1954)}]{Magnus1954}%
  \BibitemOpen
  \bibfield  {author} {\bibinfo {author} {\bibfnamefont {W.}~\bibnamefont
  {Magnus}},\ }\href@noop {} {\bibfield  {journal} {\bibinfo  {journal}
  {Communication on Pure and Applied Mathematics}\ }\textbf {\bibinfo {volume}
  {7}},\ \bibinfo {pages} {649} (\bibinfo {year} {1954})}\BibitemShut {NoStop}%
\bibitem [{\citenamefont {Moler}\ and\ \citenamefont
  {Loan}(2003)}]{19dubiousways}%
  \BibitemOpen
  \bibfield  {author} {\bibinfo {author} {\bibfnamefont {C.}~\bibnamefont
  {Moler}}\ and\ \bibinfo {author} {\bibfnamefont {C.~V.}\ \bibnamefont
  {Loan}},\ }\href@noop {} {\bibfield  {journal} {\bibinfo  {journal} {SIAM
  Review}\ }\textbf {\bibinfo {volume} {45}},\ \bibinfo {pages} {1} (\bibinfo
  {year} {2003})}\BibitemShut {NoStop}%
\bibitem [{\citenamefont {Tarantola}(2009)}]{Tarantola}%
  \BibitemOpen
  \bibfield  {author} {\bibinfo {author} {\bibfnamefont {A.}~\bibnamefont
  {Tarantola}},\ }\href@noop {} {\emph {\bibinfo {title} {Elements for
  Physics}}},\ \bibinfo {edition} {2nd}\ ed.\ (\bibinfo  {publisher}
  {Springer},\ \bibinfo {year} {2009})\BibitemShut {NoStop}%
\end{thebibliography}%

\end{document}